\documentstyle[preprint,aps]{revtex}
\newcommand{\identity}{{\sf 1 \hspace{-0.15em}
            \rule{0.087em}{1.5ex}
            \rule{0.12em}{0.1ex}
            \hspace{-0.3em}
            \rule[1.4ex]{0.12em}{0.1ex}
            \hspace{0.3em}}}
\newcommand{\bff}[1]{{\mbox{\boldmath $#1$}}}

\begin{document}
\title{Effective DBHF Method for Asymmetric Nuclear Matter and Finite Nuclei}
\author{ Zhong-yu Ma\thanks{Also Center of Nuclear Theoretical Physics,
National Laboratory of Heavy Ion Accelerator and China Institute
of Theoretical Physics.} and Ling Liu}
\address{
  China Institute of Atomic Energy, Beijing 102413  }
\date{May 5, 2002}
\maketitle

\begin{abstract}
A new decomposition of the Dirac structure of nucleon
self-energies in the Dirac Brueckner-Hartree-Fock (DBHF) approach
is adopted to investigate the equation of state for asymmetric
nuclear matter. The effective coupling constants of $\sigma $,
$\omega $, $\delta $ and $\rho $ mesons with a density dependence
in the relativistic mean field approach are deduced by reproducing
the nucleon self-energy resulting from the DBHF at each density
for symmetric and asymmetric nuclear matter. With these couplings
the properties of finite nuclei are investigated. The agreement of
charge radii and binding energies of finite nuclei with the
experimental data are improved simultaneously in comparison with
the projection method. It seems that the properties of finite
nuclei are sensitive to the scheme used for the DBHF self-energy
extraction. We may conclude that the properties of the asymmetric
nuclear matter and finite nuclei could be well described by the
new decomposition approach of the G matrix.
\end{abstract}


\section{Introduction}
The Dirac Brueckner Hartree-Fock (DBHF) approximation is a
microscopic approach to study the saturation property of infinite
nuclear matter, which is based on realistic nucleon-nucleon
interaction and contains features of the relativistic
theory\cite{Machl89,Brock90,Jong980,Muther90}. The presently
popular relativistic mean field (RMF) theory\cite{Serot86,Ring96}
is considered to be a phenomenological method in comparison with
the DBHF approach. There are at least six free parameters in the
RMF calculations, such as NL1\cite{Rufa86}, NL3\cite{Lala97} and
NL-SH\cite{Sharma93}. These parameters are adjusted by reproducing
nuclear matter saturation properties and bulk properties of large
number of stable nuclei, such as binding energies and
charge radii. Consistent results with various parameter sets can
be achieved for stable nuclei. However, discrepancies occur when
the calculation is extended to nuclei in drip lines or super-heavy
nuclei. The DBHF method adopts the realistic nucleon-nucleon (NN)
interaction which is fitted to the NN scattering phase shifts and
deuteron properties. It takes the nucleon in-medium short-range
correlation effect into account by performing a complete summation
of two-body ladder diagrams. Therefore the DBHF is generally
accepted as one of the most reliable and feasible microscopic
method for the description of effective interactions in the
nuclear medium. A break-through with the DBHF theory was that it
reproduces empirical saturation properties of nuclear matter
successfully\cite{Brock90}, and presents a reasonable nuclear
equation of state (EOS). This kind of EOS, obtained directly from a
microscopic theory, contains more physical meanings than
phenomenological models, especially for the investigation of the
high-density and high-temperature behavior in the astrophysics as
well as the collision of energetic heavy ions.

Attempts have been made to calculate the finite nuclei with
effective meson-nucleon interactions, which are deduced from the
DBHF self-energy. One of the most successful approaches in this
direction is the so called relativistic density dependent
Hartree\cite{Brock92} or Hartree-Fock\cite{Zhongy94,Hual95}, which
could reasonably describe the properties of stable nuclei. In this
approach the coupling constants of isoscalar $\sigma$ and $\omega$
mesons are adjusted at each density  by reproducing the nucleon
self-energies resulting from the DBHF. In this method, the Dirac
structure of the self-energy in the DBHF G matrix was extracted
from the momentum dependence of the single particle energy, where
the momentum dependence of scalar and vector potentials was
neglected\cite{Brock90}. This simple method encountered an
obstacle in extracting the effective coupling constants of
isovector mesons when it extends to asymmetric nuclear
matter\cite{Shen97}. It was discussed that such approach may yield
an isospin dependence of the effective interaction with a wrong
sign\cite{Ulrych97}. Another scheme to extract the Dirac energy is
the so called projection method\cite{Boersma94,Jong98}. Some
ambiguities with pseudo-scalar and pseudo-vector couplings may be
observed in the projection method due to the fact that positive
energy Dirac spinors are used in the DBHF calculation only. In
order to overcome these drawbacks, recently Schiller and
Muether\cite{Schiller01,Muet01} suggested a new decomposition
approach of the DBHF G-matrix in extracting the Dirac structure of
the nucleon self energy. They separated the G-matrix into the bare
NN interaction V and correlation $\Delta G$. The projection
method is applied only on the correlation term $\Delta G$, which
is parameterized by four pseudo-mesons. The
ambiguities in the projection method are eliminated and a
satisfactory description for the symmetric nuclear matter is
achieved in this scheme.

In this paper we shall study the properties of asymmetric
nuclear matter as well as neutron matter with the new
decomposition of the Dirac structure in the DBHF G matrix. The
main purpose of this paper is to extract an effective NN
interaction with an isospin dependence from the DBHF nucleon
self-energy in the symmetric and asymmetric nuclear matter. The
effective nucleon-meson coupling constants with a density
dependence contain the DBHF short-range correlations and
information of the isospin structure. The properties of finite
nuclei are investigated with these effective coupling constants.

 This paper is arranged as the follows. In Sec.II the new
decomposition of the Dirac structure of the DBHF G matrix is
presented. The properties of asymmetric nuclear matter are
discussed in Sec.III. The effective NN interaction with density
dependent nucleon-meson couplings is extracted from the DBHF G
matrix, which is described in Sec.IV. Studies of finite nuclei
with the effective NN interaction in the RMF are depicted in
Sec.V. Finally we give a brief summary in Sec.VI.

\section{A new decomposition of DBHF G matrix}

In the new decomposition approach, the DBHF G matrix is split into
two parts\cite{Schiller01}, the bare NN interaction $V$ and
correlation term $\Delta G$ :
\begin{equation}
G=V+\Delta G~~.  \label{eq1}
\end{equation}
Usually the bare NN interaction $V$ in the DBHF calculation is
taken as one boson exchange potential (OBEP), such as Bonn
potentials. The OBEP of Bonn potentials is associated with three
isoscalar mesons ($\sigma $, $\omega $ and $\eta $) and three
isovector mesons ($\delta $, $\rho $ and $\pi $) with the
following quantum numbers ($J^{\pi }$, $T$): $\sigma
(0^{+},0),~~\omega (1^{-},0),~~\eta (0^{-},0),~~\delta
(0^{+},1),~~\rho (1^{-},1),~~\pi (0^{-},1)$. The Dirac structure
of the OBEP is explicitly known and its contribution to the
nucleon self-energy can be calculated in the relativistic
Hartree-Fock approach. Analyzing the correlation term $\Delta G$,
the peculiar property of the dominating $\pi $-exchange
contribution is removed, therefore the projection method can be
applied to $\Delta G$ without ambiguities. Schiller and
Muether\cite{Schiller01,Muet01} observed that the inclusion of
$\Delta G$ leads to a constant shift in the nucleon scalar and
vector potentials in comparison with those calculated with only
OBEP, which is almost independent of the nucleon momentum.
Therefore, the correlation term $\Delta G$, can be well described
in terms of an effective interaction with zero range. It is
parameterized by four isoscalar and isovector effective mesons :
($\sigma ^{\prime }$, $\omega ^{\prime }$, $\delta ^{\prime }$ and
$\rho ^{\prime }$) with infinite masses, which are defined as
pseudo-mesons. The ratios of coupling constants and corresponding
masses for those psuedo-mesons are finite and of weak density
dependence. In this way, the contribution of the correlation
effect $\Delta G$ to the self energy can also be calculated in the
relativistic Hartree-Fock approach.

The self-energy of protons and neutrons with a momentum $k$ in
nuclear matter can be decomposed into scalar and time-like and
space-like vector components.
\begin{equation}\label{eq2}
  \Sigma^i(k)=\Sigma_s^i(k)- \gamma^0\Sigma_0^i(k) + {\bff \gamma} \cdot
  {\bf k} \Sigma_v^i(k)~~,
\end{equation}
where $i$ denotes the proton and neutron. It is convenient to
eliminate the space-like vector component $\Sigma _{v}^{i}$ and
write the Dirac equation into a form which  contains only scalar
and time-like vector components in a usual way,
\begin{equation}
\left\{ \vec{\gamma}\cdot \vec{k}+[M+
\widetilde{\Sigma}_{s}^{i}(k)]-\widetilde{\Sigma}_{0}^{i}(k)\gamma
^{0}\right\} u_{i}(k)=\varepsilon _{i}(k)\gamma ^{0}u_{i}(k)~,
\label{eq3}
\end{equation}
where
\begin{equation}\label{4}
  \widetilde{\Sigma}_s^i=\frac{\Sigma_s^i-M\Sigma_v^i}{1+\Sigma_v^i}, ~~~~~~~~
  \widetilde{\Sigma}_0^i=\frac{\Sigma_s^i-\varepsilon_i \Sigma_v^i}{1+\Sigma_v^i}~.
\end{equation}
 We define the nucleon effective mass,
\begin{equation}
M^{\ast }(k)=M+ \widetilde{\Sigma}_{s}^{i}(k)~. \label{eq5}
\end{equation}
The single-particle energy takes the form,
\begin{equation}
k_{i}^{0}=\varepsilon _{i}(k)=\sqrt{k^{2}+M_{i}^{\ast }(k)^{2}}-
\widetilde{ \Sigma}_{0}(k)=E^{\ast }(k)-
\widetilde{\Sigma}_{0}(k)~. \label{eq6}
\end{equation}

The Dirac structure of the nucleon self-energy in nuclear matter
in the DBHF approach can be calculated with the OBEP $V$ and
$\Delta G$, which is parameterized with psuedo-mesons.
\begin{equation}
\Sigma (k)=\sum_{\alpha }\int \frac{d^{4}q}{(2\pi )^{4}}\{\Gamma
_{\alpha }^{a}D_{\alpha }^{ab}(0)Tr[i{\cal{G}}(q)\Gamma _{\alpha
}^{b}]-\Gamma _{\alpha }^{a}[i{\cal{G}}(q)]D_{\alpha
}^{ab}(k-q)\Gamma _{\alpha }^{b}\}~~,  \label{eq7}
\end{equation}
where the first and second terms are direct and exchange
contributions, respectively. The index $\alpha $ refers to mesons
and $a$, and $b$ refer to isospin components. ${\cal{G}}(q)$ is the
single particle Green's function,
\begin{equation}\label{eq7.5}
  {\cal{G}}(k)=(\gamma^\mu k_\mu +M^*) \left(
  \frac{1}{k_\mu^2-M^{*2}+i\eta} + \frac{i\pi}{\varepsilon_k}
  \delta(k_0-\varepsilon_k)\Theta(k_F-|{\bff{k}}|) \right)~~.
\end{equation}
$D_{\alpha }^{ab}(q) $ are meson propagators, which  have the
following forms for scalar and vector mesons, respectively.
\begin{eqnarray} \label{eq8}
D_s^{ab}(k) &=&\frac 1 {k_\mu^2-m_s^2+i\eta } T^{ab}  \\
D_v^{ab}(k) &=&\frac{-g_{\mu \nu }+k_\mu k_\nu
/m_v^2}{k_\mu^2-m_v^2+i\eta } T^{ab}, \nonumber
\end{eqnarray}
where $T^{ab}$ is the isospin operator, $\identity$ for isoscalar
and  $\delta_{ab}$ for isovector mesons, respectively. The vertex
operators $\Gamma _{\alpha }^{a}$ for scalar, vector and
pseudoscalar mesons are expressed as the follows. The vector
mesons have both vector and tensor couplings and  pseudo-vector
couplings for pseudoscalar mesons are adopted.
\begin{equation}
\Gamma _{\alpha }^{a}=t^{a}\left\{
\begin{array}{c}
ig_{s} ~~~~ \text{ \qquad for scalar mesons} \\
i(g_{v}\gamma ^{\mu }+\frac{f_{v}}{2M}\partial _{\nu }\sigma ^{\nu
\mu })~~~~
\text{ \qquad  for vector mesons} \\
i\frac{f_{ps}}{m_{ps}}\gamma ^{5}\gamma ^{\mu }\partial _{\mu }\text{ \qquad
for pseudo-scalar mesons}
\end{array}
\right.  \label{eq9}
\end{equation}
The isospin operator $t^{a}$ equals $\identity $  for isoscalar
mesons and $\tau _{a}$ for isovector mesons.

\section{Asymmetric nuclear matter}

For asymmetric nuclear matter one defines the asymmetry
parameter, $\beta =(\rho ^{n}-\rho ^{p})/(\rho ^{p}+\rho ^{n})$,
where $\rho ^{p}$ and $\rho ^{n} $ denote the proton and neutron
density, respectively. It implies that $\beta =0$ for the
symmetric nuclear matter and $\beta =1$ for the pure neutron
matter. The nucleon self-energies for protons and neutrons have to
be calculated individually.

The scalar and vector potentials in symmetric nuclear and neutron
matter with the $G$ matrix have been investigated in
Ref.\cite{Schiller01,Liu02}. Significant differences in scalar and
vector potentials were observed in comparison with those obtained
by using a simple method, where the nucleon self-energy was
deduced from the single-particle energy, although the EOS produced
by both methods were very similar. In this paper we present the
study of the EOS for asymmetric nuclear matters. The Bonn A
potential is chosen as the bare NN potential in the DBHF
calculation and the coupling constants of pseudo mesons are taken
from Ref.\cite{Schiller01}. The EOS for various asymmetric
parameters are given in Fig.1. For the asymmetry parameter $\beta
< 0.8$, the energy per nucleon as a function of the density
exhibits a curve with a minimum, but the saturation point shifts
to lower densities as the asymmetry parameter increases. However,
for the extreme neutron-rich or neutron matter ($\beta =1$) no
minimum in the EOS is found.

The EOS of nuclear and neutron matters are two extreme cases,
which determine the density dependence of the asymmetry energy
$a_{asym}(\rho )$. In Fig.2(a) bounding energies per nucleon
$E/A(\rho,\beta)$ at various densities are plotted as a function
of $\beta^2$. Straight lines are drawn from the linear fit of the
first three points at small asymmetric parameters $\beta$. It is
found that those lines go through most numerical points and only
small deviations are observed at $\beta =1 $. It illustrates that
the empirical parabolic law is fulfilled, which is consistent with
the finding in the non-relativistic approach\cite{Zuo99}.
Therefore the asymmetry energy $a_{asym}(\rho )$ could be
expressed as $E/A(\rho ,\beta =1)-E/A(\rho ,\beta =0)$. The
density dependence of the asymmetric energy in nuclear matter is
shown in Fig.2(b). The dashed curve is obtained in a
non-relativistic BHF calculation taken from Ref.\cite{Bom91}. Both
of them give similar asymmetric energy around the saturation
density and below, but diverge at large densities.

\section{Effective NN interaction}

From scalar and vector self-energies in asymmetric nuclear matter
calculated with the effective G matrix we could extract the
density dependence of effective coupling constants of isoscalar
and isovector mesons. Following Brockman's method\cite{Brock92} we
impose a condition that the nucleon self-energy at each nuclear
density and asymmetric parameter obtained in the DBHF is
reproduced by the effective NN interaction in the RMF calculation.
The momentum dependence of the self-energy $
\widetilde{\Sigma}_s^i$ and $ \widetilde{\Sigma}_0^i$ is rather
weak, so it is reasonable to take their mean values $U_s^i$   and
$U_0^i$   from $k=0$ to $k=k_F^i$. The effective coupling
constants of $\sigma$, $\omega$, $\delta $ and $\rho $ mesons can
be determined as the follows,
\begin{eqnarray}
\left( \frac{g_{\sigma }(\rho_B)}{m_{\sigma }}\right) ^{2} &=&-\frac{1}{2}%
\frac{U_{s}^{p}(\rho_B)+U_{s}^{n}(\rho_B)}{\rho _{s}^{p}+\rho
_{s}^{n}}  \label{eq10} \\
\left( \frac{g_{\delta }(\rho_B)}{m_{\delta }}\right) ^{2} &=&-\frac{1}{2}%
\frac{U_{s}^{p}(\rho_B)-U_{0}^{n}(\rho_B)}{\rho_s  ^{p}-\rho_s
 ^{n}}  \nonumber \\
\left( \frac{g_{\omega }(\rho_B)}{m_{\omega }}\right) ^{2} &=&-\frac{1}{2}%
\frac{U_{0}^{p}(\rho_B)+U_{0}^{n)}(\rho_B)}{\rho ^{p}+ \rho^{n}}  \nonumber \\
\left( \frac{g_{\rho }(\rho_B)}{m_{\rho }}\right) ^{2}
&=&-\frac{1}{2} \frac{U_{0}^{p}(\rho_B)-U_{0}^{n}(\rho_B)}{\rho
^{p}-\rho ^{n}}~, \nonumber
\end{eqnarray}
where $\rho_B=\rho^p + \rho^n$. In general, the ratios of $g/m$
could be functions of the scalar or vector density and both the
coupling constant and mass are of density dependence. In this
paper we assume that only coupling constants depend on the vector
density $\rho_B$ and masses of mesons remain constant. The
coupling constants of $\sigma $ and $\omega $ mesons are taken
from their values at $\beta =0$ in order to reproduce the DBHF
results in nuclear matter. For the isospin dependence, we
calculate the coupling constants for isovector mesons in various
asymmetric nuclear matters with $\beta$ = 0.2, 0.4, and 0.6,
respectively, and then take their averaged values.

The coupling constants as a function of the density are displayed
in Fig.3. The solid circles are the results obtained from the
nucleon self-energies in Eq.\ref{eq10}. Due to a phase transition
at very low density $ k_F<1.0$ fm$^{-1}$,  the DBHF calculation can
not be applied. Therefore, the extrapolation of coupling
constants at low densities has to be performed when one tries to
study the properties of finite nuclei. To give a guide line of the
extrapolation by fittings, the Hartree-Fock calculation with the G
matrix at $k_F$ = 1.0 fm$^{-1}$ is carried out, which are plotted
by open circles. The solid curves are drawn by a fitting with
exponential forms as : $ g_\alpha^2(\rho _B)=a_0+a_1\exp (-\rho
_B/a_2)$  for $\alpha=\sigma ,\omega ,\rho$ and a polynomial form:
$g_\delta^2 (\rho _B)=a_0+a_1\rho _B+a_2\rho _B^2$ for $\delta$
meson. The parameters of fitting are listed in Table. 1 and marked
by Fit.A.

The RMF calculation by the effective nucleon-meson couplings
reproduces the DBHF nucleon self-energies, but not the EOS. The
deviation of the binding energy at the saturation point is within
2\% and increases at high densities. We also perform another
extraction procedure for effective couplings by imposing the
condition of reproducing the DBHF scalar potential and binding
energy per nucleon at each density for symmetric and asymmetric
nuclear matters. In this way, the EOS and asymmetry energy in the
DBHF calculation are exactly reproduced in the RMF calculation
with the new set of effective nucleon-meson couplings. The
parameters of this kind of fitting are also shown in Table. 1 and
denoted by Fit.B.

In Fig.3 we plot the coupling constants obtained from
Ref.\cite{Brock90,Brock92} and Ref.\cite{Hofm01} for a comparison.
The dashed curves are the results in Ref.\cite{Brock90,Brock92},
where the Dirac structure of nucleon self energies in the DBHF
calculation is extracted from the momentum dependence of the
single particle energy. These coupling constants are obtained
through those scalar and vector self-energies at nuclear and
neutron matter. It is found that the results are very different
from ours, although the DBHF G matrix is obtained from the same
OBEP of Bonn A in both methods. The dependence of coupling
constants on the density in Ref.\cite{Shen97} is extremely strong.
Furthermore, the absolute values of isovector mesons ($g_\delta $,
$g_\rho $) are very large and have the same magnitude as those for
isoscalar mesons ($g_\sigma $, $g_\omega $). It should be noted
that the square values of coupling constants for isovector mesons
turn out to be negative in that analysis, which means that the
coupling constants themselves would be imaginary. Those are not
consistent with our common understanding. It can be seen from
those solid curves that our coupling constants exhibit a moderate
density dependence, and the strengths for isovector mesons are
much weaker than those of isoscalar mesons. The dotted curves in
Fig.3 are the results taken from Ref.\cite{Hofm01}, where the DBHF
self-energy is extracted in terms of a projection method. Those
couplings are taken from their average values obtained in
asymmetry nuclear matters with $\beta$ = 0.2, 0.4, and 0.6. It is also
seen that the coupling constants from the projection method is
stronger than our results, especially for isovector mesons.

\section{Finite nuclei}

We study the properties of some doubly magic and semi-magic
nuclei, such as $^{16}$O, $^{40}$Ca, $^{48}$Ca, $^{90}$Zr,
$^{208}$Pb, $^{48}$Ni, $^{56}$Ni, $^{68}$Ni, $^{100}$Sn, and
$^{132}$Sn in the RMF with the effective NN interaction. The
rearrangement terms due to the density dependence in the effective
nucleon-meson couplings are included in the calculation. The
relative deviations of charge radii and binding energies
$(x_{Cal.}-x_{Exp.})/x_{Exp.}$ are displayed in Figs.4(a) and
4(b), respectively. The solid (open) circles are our results
calculated with the effective nucleon-meson couplings of Fit.A
(Fit.B). The calculated charge radii are improved in the case of
Fit.B, but the resulting binding energies become smaller in
comparison with the experiment data. In a general speaking, an
increase in the charge radius produced by one set of parameters
would result in a decrease of the value of the binding energy. A
better way to estimate a scheme is to investigate whether the
calculated binding energies and charge radii could both be
improved. In Ref.\cite {Hofm01}, they adopted a momentum
correction in the extraction of couplings from the DBHF self
energies in a projection method. In this way additional parameters
are introduced, which are adjusted to the nuclear matter or finite
nuclei. With the momentum correction they can improve the binding
energy and charge radii at the same time. The upper and lower
triangles are the results in Ref.\cite{Hofm01} corresponding to
the cases adjusted to the nuclear matter and finite nuclei,
respectively. In our calculation we do not introduce additional
parameters, except for the extrapolation at low densities, which
are slightly adjusted by the properties of finite nuclei. Our
results are improved simultaneously in binding energies and charge
radii in comparison with those from Ref.\cite{Hofm01}, although
the G matrix in Ref.\cite{Hofm01} have been calculated from the
Groningen parameterization. Actually the saturation properties in
nuclear matter from the DBHF calculation with those two free NN
interactions : Bonn A and Groningen interaction are very similar,
which are $E/A$ = -15.6 MeV at $\rho _{B0}$ = 0.185 $fm^{-3}$ and
$E/A$ = -15.5 MeV at $\rho _{B0}$ = 0.182 $fm^{-3}$, respectively.
However, both of binding energies and charge radii of our results
are better than those in Ref.\cite{Hofm01}.  It seems that the
properties of finite nuclei are sensitive to the scheme used for the
DBHF self energy extraction.

The neutron and proton spin-orbit splittings
$E_{SO}=E_{n,l,j=l-1/2}-E_{n,l,j=l+1/2}$ are examined in Fig.5 for
$1p$ shell for $^{16}$O and the $1d$ shell for $^{40}$Ca,
$^{48}$Ca, and $^{48}$Ni. The solid and open circles represent the
calculations with couplings of Fit.A and Fit.B in Table 1,
respectively. Various experimental data\cite{Fuck95,Lo00} are also
shown in Fig.5. Although the experimental values are not
accurately established, our results are in good agreement with the
experimental data except for $^{48}$Ca. The experimental data for
$^{40}$Ca and $^{48}$Ca are divergent. The results show that the
effective interaction deduced from the new Dirac structure
decomposition of the G matrix could well describe the properties
of finite nuclei in comparison with other methods.

\section{A brief summary}

In summary, we have applied the new decomposition of the Dirac
structure of the nucleon self energy from the G matrix in the DBHF
calculation to investigate properties of symmetric and
asymmetric nuclear matter. The DBHF G matrix is composed of a
OBEP and correlation term, which are parameterized by four
pseudo-mesons with infinite masses. The EOS for asymmetric nuclear
matter and the density dependence of the asymmetry energy are
investigated with the DBHF G matrix. The effective NN interaction
with a density dependence in the RMF is extracted from the DBHF G matrix. The
effective coupling constants of isoscalar and isovector mesons
incorporate the DBHF results and involve NN correlation effects.
We performed calculations for finite nuclei using the effective
nucleon-meson coupling constants. The bulk properties of finite
nuclei are well described and the results are in good agreement
with the experimental data. Both of binding energies and charge
radii are improved simultaneously in comparison with the
others\cite{Hofm01}, the deviations of which with the experimental
data for some magic and semi-magic nuclei are within 5\%. It seems
that the properties of finite nuclei are sensitive to the scheme
used for the DBHF self energy extraction. Examinations of the
spin-orbit split are also made. The results show good
agreement with the experimental data available. Finally we may
conclude that the splitting of the G matrix into a bare interaction
and correlation term is a satisfactory way to extract the Dirac
structure of the nucleon self energy in the DBHF. The properties
of asymmetric nuclear matter and finite nuclei could be well
described by the new decomposition approach of the G matrix. The
application of this approach to the neutron star and investigation
of effects of isovector mesons in exotic nuclei are the future
interesting exploration.

{\bf ACKNOWLEDGMENTS}

One of the authors MZY would like to thank Prof. H. Muether and
Dr. E. Schiller for stimulating discussions and providing the
parameters of the Dirac structure of the G matrix. This work is
supported by the National Natural Science Foundation of China
under No. 10075080, 19835010 and Major State Basic Research
Development Program under contract No. G20007740.

\begin{table}
\begin{tabular}{p{1.4in}p{1.2in}p{1.2in}p{1.2in}}
   & $\quad a_0$ & $a_1$ & $a_2$ \\ \hline
Fit.A $ \qquad g_\sigma^2 $ & 76.8226 & 48.4004 & 0.1332 \\
$\qquad \qquad g_\omega^2 $ & 97.7896 & 71.9705 & 0.2387 \\
$\qquad \qquad g_\rho^2 $ & 11.5862 & 21.3433 & 0.1385 \\
$\qquad \qquad g_\delta^2 $ & 5.4139 & 34.7149 & 15.4454 \\
\hline
Fit.B $  \qquad g_\sigma^2 $ & 77.7312 & 48.5497 & 0.1260 \\
$\qquad \qquad g_\omega^2 $ & 104.8553 & 69.0250 & 0.1955 \\
$\qquad \qquad g_\rho^2 $ & 10.8791 & 34.1492 & 0.0845 \\
$\qquad \qquad g_\delta^2 $ & 7.1379 & 16.9648 & 52.2796 \\
\end{tabular}
\caption{ Parameters of effective coupling constants by fittings.
The expressions of fittings are described in the text. Fit.A and
Fi.B denote the fittings through the DBHF scalar and vector
self-energies and by the DBHF scalar and binding energies,
respectively.}
\end{table}

\begin{center} {\bf FIGURE CAPTIONS} \end{center}

\noindent FIG. 1. The equation of state for various asymmetry
parameters from symmetric nuclear matter($\beta =0)$ to neutron
matter ($\beta =1$).

\noindent FIG. 2(a). The binding energy per nucleon
$E/A(\rho,\beta)$  as a function of asymmetry parameter
$\beta^2$. Fig. 2(b). The density dependence of the asymmetry
energy. The dashed curve is the result of a non-relativistic
Brueckner Hartree-Fock calculation\cite{Zuo99}.

\noindent FIG. 3. The density
dependence of coupling constants for $\sigma$, $\omega$, $\delta $
and $\rho $ mesons. The solid circles are extracted from the DBHF
self-energies in the present work. Open circles are obtained by
extrapolation calculations with the G matrix at $k_F$ = 1.0
fm$^{-1}$. The solid curves are by fittings with an exponential or
polynomial form Fit.A in Table 1. The dashed lines are the results
in Ref.\cite{Brock90,Brock92}, where the DBHF self-energies are
determined from the single-particle energy. The dotted lines are
obtained in Ref.\cite{Hofm01} with a projection method.

\noindent FIG. 4. The relative deviations of charge radii(a) and
the binding energies(b) for some stable nuclei. The solid and open
circles are our results with coupling constants Fit.A and Fit.B in
Table.1, respectively. The upper and lower triangles are taken
from Ref.\cite{Hofm01} adjusted to the nuclear matter and finite
nuclei, respectively. Lines are drawn to guide the eyes.

\noindent FIG. 5. The spin-orbit split for the proton (a) and neutron
(b). The notations are the same as in Fig.4. The open lower
triangles are also taken from Ref.\cite{Hofm01}, which are the
same as the case with lower solid triangles except for that the
Bonn A potential was adopted. The experimental data denoted by
A\cite{Fuck95} and B, C, D and E are taken from Ref.\cite{Lo00}
and references cited therein.
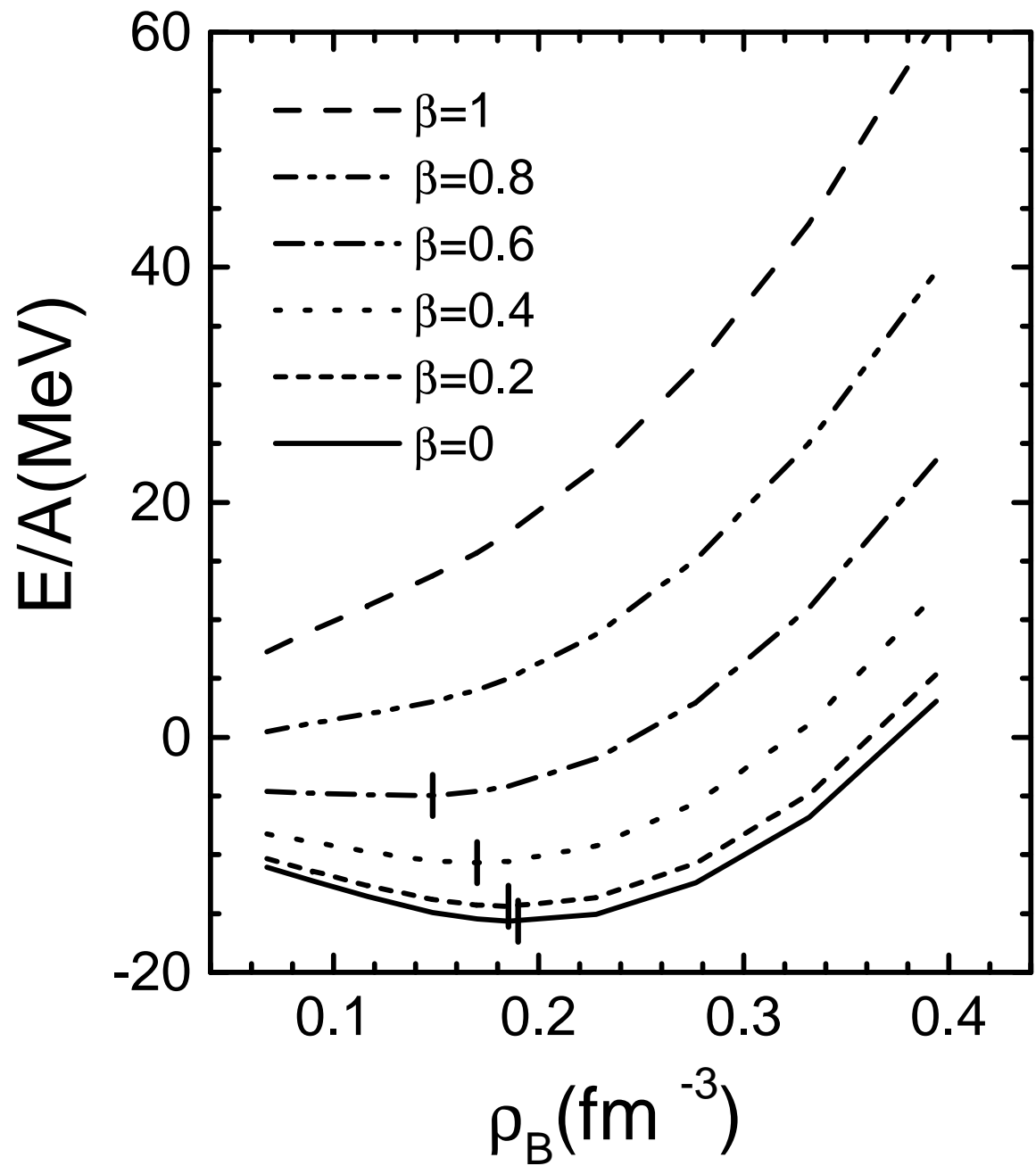
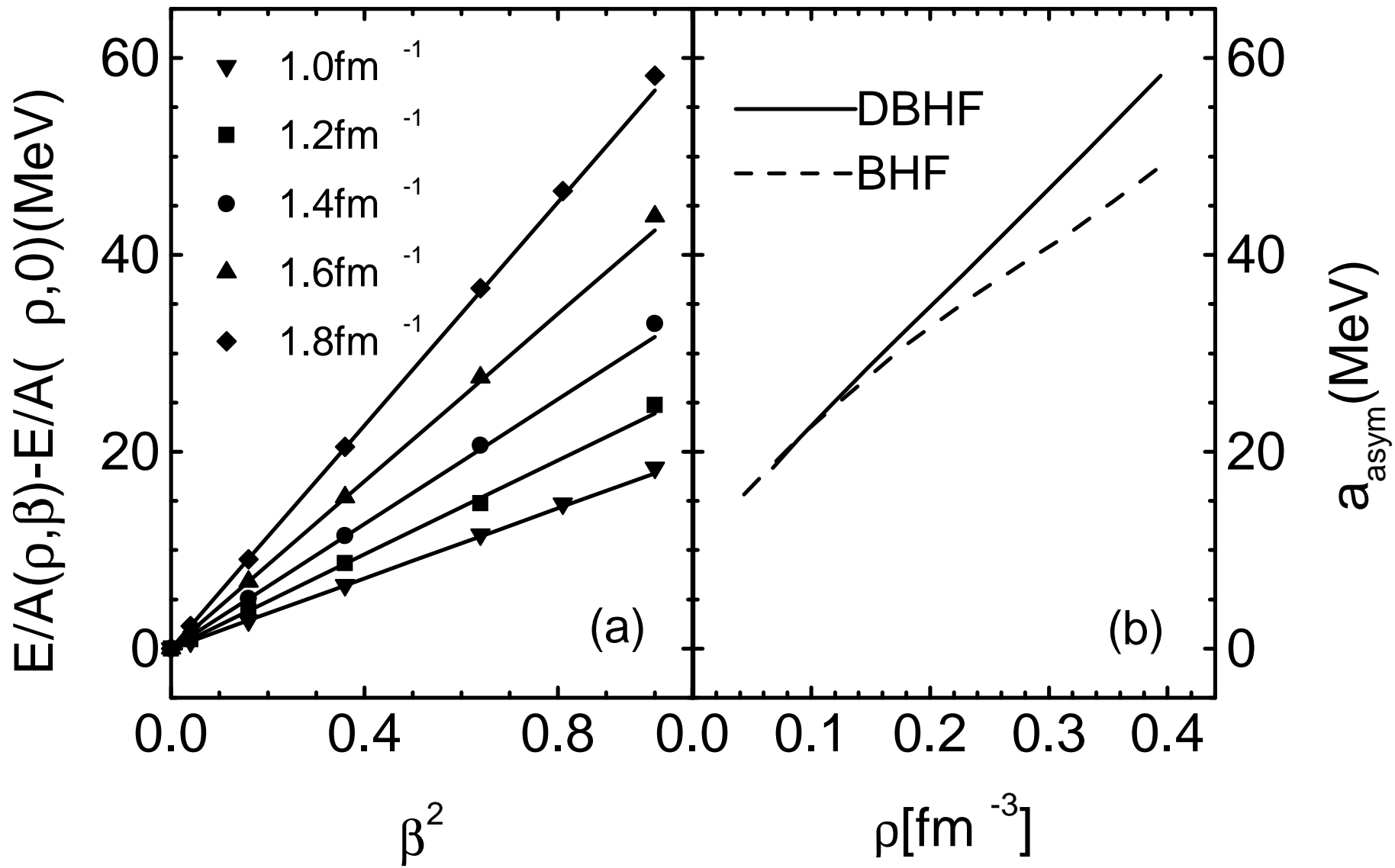
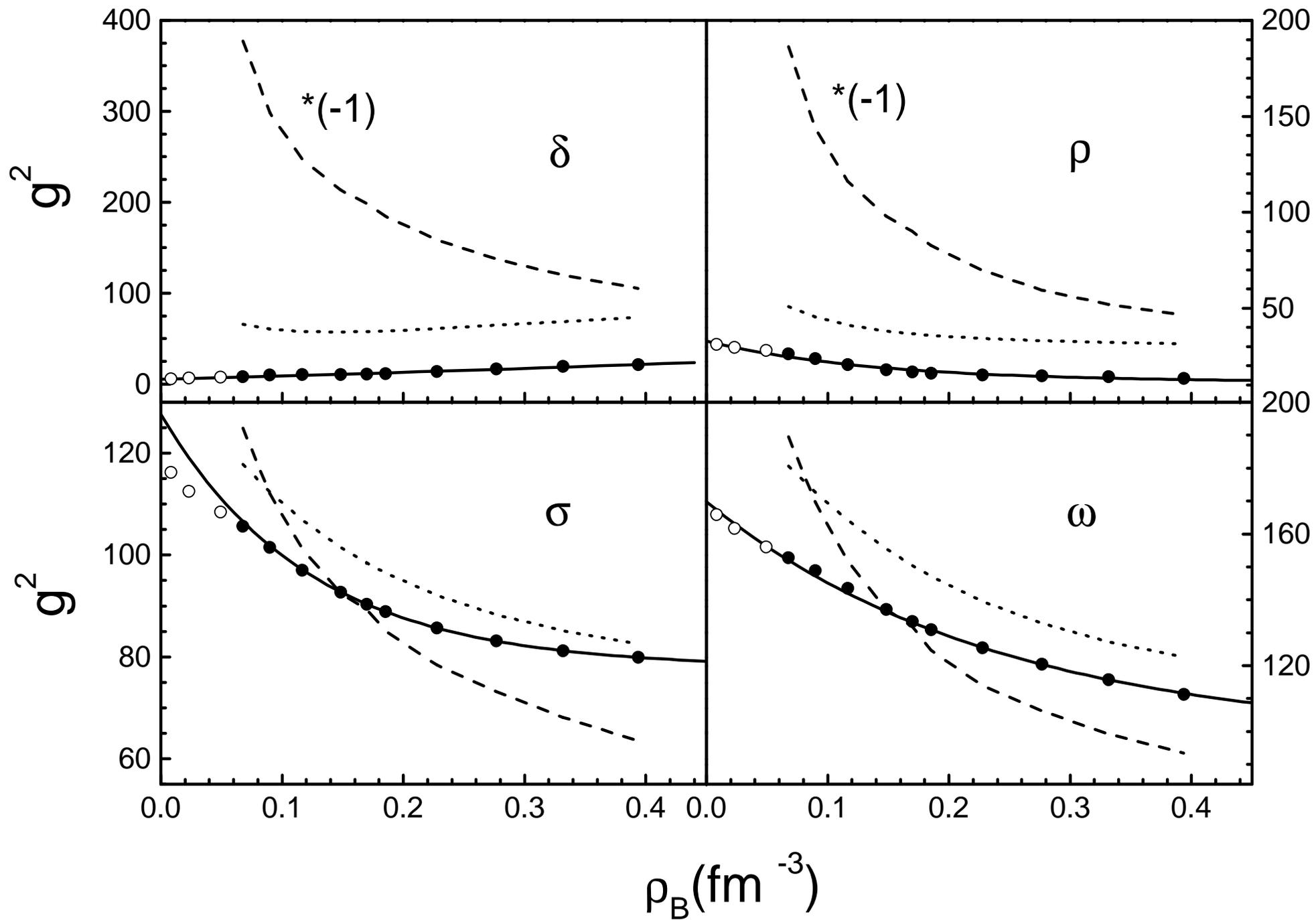
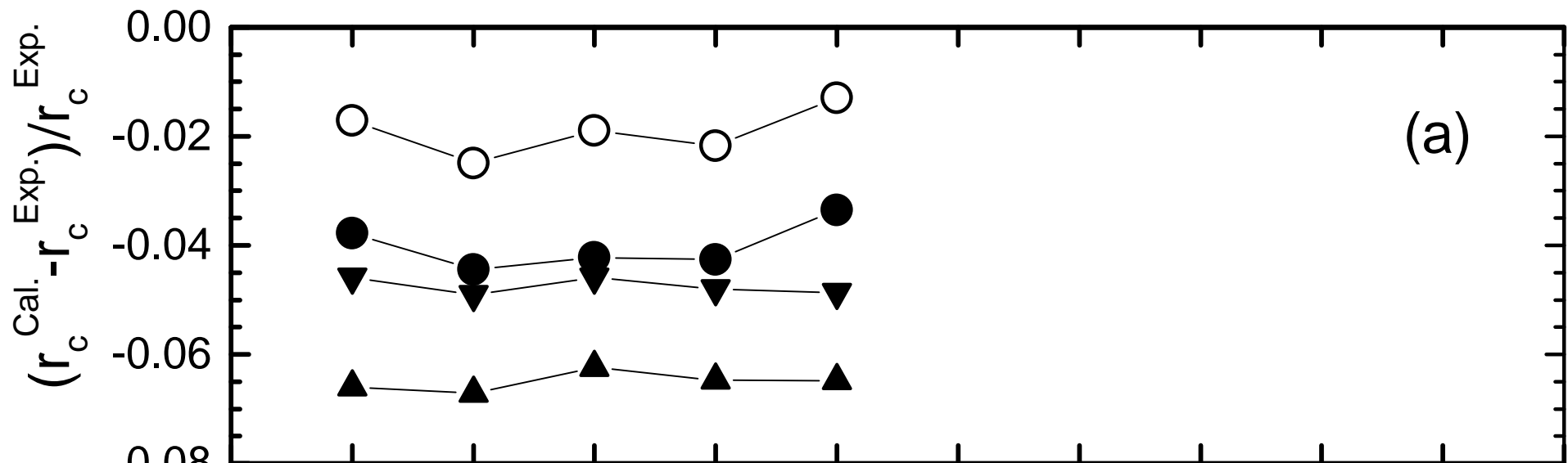
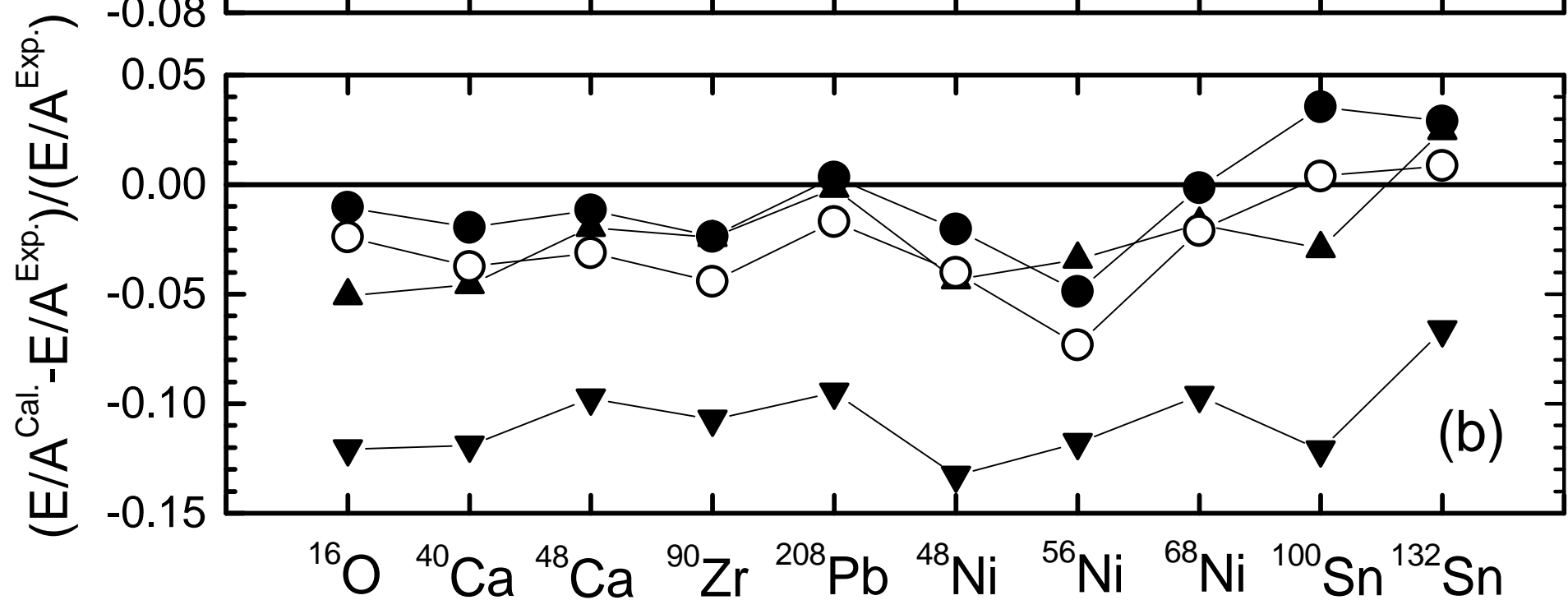
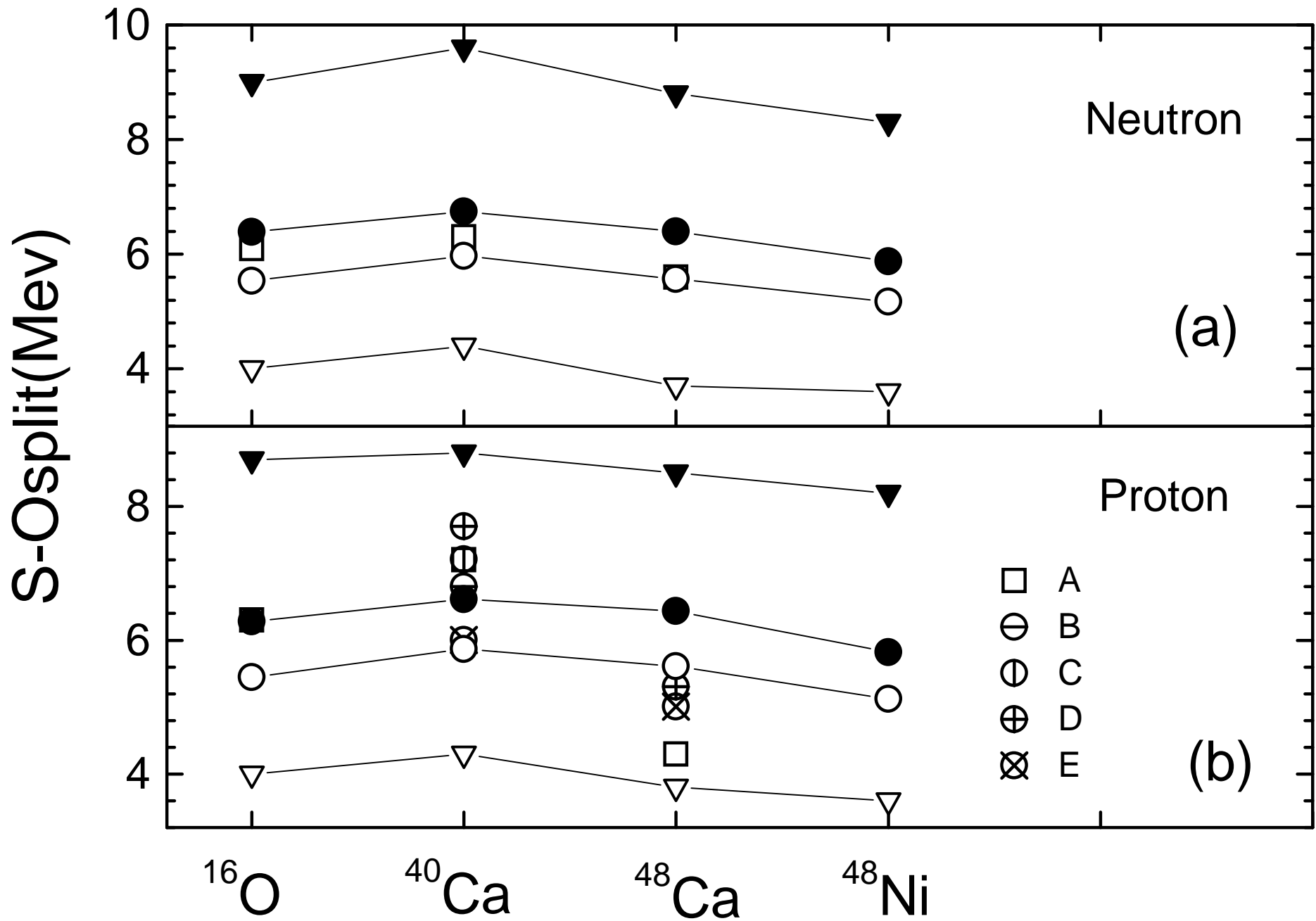
\end{document}